\documentclass[prl,twocolumn,showpacs,nobalancelastpage,amsmath,amssymb]{revtex4}
\usepackage{graphicx}
\usepackage{bm}
\usepackage{bbm}
\begin{document}
\date{\today}

\title{
Magnetoelectric Andreev effect due to proximity-induced non-unitary triplet superconductivity in helical metals 
}

\author{G. Tkachov}

\affiliation{
Institute for Theoretical Physics and Astrophysics, Wuerzburg University, Am Hubland, 97074 Wuerzburg, Germany
}

\begin{abstract}
Non-centrosymmetric superconductors exhibit the magnetoelectric effect which manifests itself in the appearance of the magnetic spin polarization in response to 
a dissipationless electric current (supercurrent). While much attention has been dedicated to the thermodynamic version of this phenomenon (Edelstein effect), 
non-equilibrium transport magnetoelectric effects have not been explored yet. We propose the magnetoelectric Andreev effect (MAE) which consists in the generation 
of spin-polarized triplet Andreev conductance by an electric supercurrent. The MAE stems from the spin polarization of the Cooper-pair condensate due to 
a supercurrent-induced non-unitary triplet pairing. We propose the realization of such non-unitary pairing and MAE in superconducting proximity structures based on 
two-dimensional helical metals -- strongly spin-orbit-coupled electronic systems with the Dirac spectrum such as the topological surface states.  
Our results uncover an unexplored route towards electrically controlled superconducting spintronics and 
are a smoking gun for induced unconventional superconductivity in spin-orbit-coupled materials. 
\end{abstract}

\maketitle

{\em Introduction.} 
Spin-triplet superconductivity bears rich many-body physics \cite{Leggett75,Sigrist91,Mackenzie03,Bergeret05,Kastening06,Tanaka12}, 
building the basis for the emerging synergy between superconducting and spin electronics \cite{Maekawa06,Eschrig11,Linder15,Eschrig15,Anwar16}. 
Unlike its single-particle counterpart \cite{Zutic04}, superconducting spintronics relies on the total spin $S=1$
of triplet Cooper pairs to manipulate electric currents. At present, much effort has been focused on magnetic control of triplet currents 
in superconducting spin valves (see review in Ref. \cite{Linder15}). 
On the other hand, for creating superconducting spintronic circuits it would be highly desirable 
to be able to manipulate triplet pairing by electric fields or currents.
While this has been recognized as an important outstanding problem, the only mechanism of the electrically tunable Cooper pairing that has been identified so far 
is the electric gating mediated by spin-orbit coupling (SOC) (see, e.g., Refs. \cite{Yada09,Ouassou16}). 

\begin{figure}[t]
\includegraphics[width=85mm]{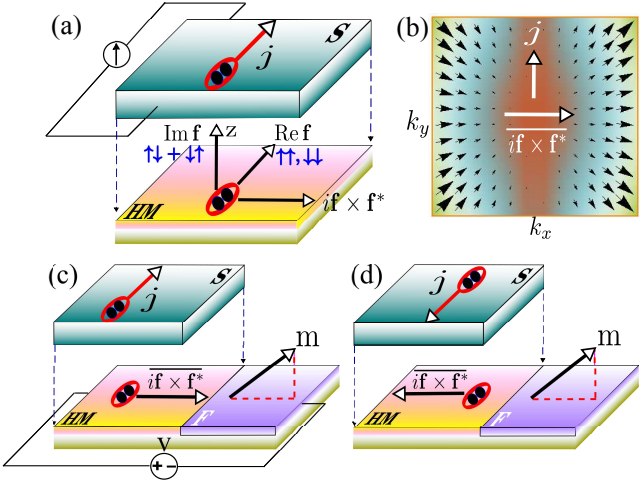}
\caption{
(Color online) 
(a) Schematic of a helical metal (HM) proximitized by a current-biased superconductor (S); ${\bm j}$ is the supercurrent density.
The supercurrent generates non-unitary triplet pairing in the HM described by the complex pair amplitude ${\bm f}$ 
with orthogonal real and imaginary parts [see also Eq. (\ref{f_Ek_2})].  
(b) Vector plot of the Cooper-pair spin polarization (CSP) $i {\bm f}(0,{\bm k},{\bm q}) \times {\bm f}^*(0,{\bm k},{\bm q})$ in wave-vector space for Dirac fermions. 
The CSP magnitude scales with the size of the black arrows. $\overline{i {\bm f}\times {\bm f}^*}$ shows the average CSP [see also Eq. (\ref{ff*_Dirac})].     
(c) and (d) Supercurrent-operated S/HM/ferromagnet (F) structure to realize the magnetoelectric Andreev effect. 
}
\label{f_j}
\end{figure}

In this Letter we explore another possibility inspired by the magnetoelectric (Edelstein) effect in superconductors with Rashba SOC \cite{Edelstein95}.
This effect consists in the generation of the thermodynamic spin magnetization by an electric supercurrent \cite{Edelstein95,Yip02,Bauer12,Konschelle15}.
We propose that the same mechanism can be used to generate spin-polarized (non-unitary) triplet pairing in two-dimensional helical metals -- 
SO-coupled electronic systems with the Dirac spectrum such as the surface states of topological insulators \cite{Hasan10,Qi11,Franz13}. 
In non-unitary triplet superconductors \cite{Leggett75,Sigrist91}, Cooper pairs carry 
an intrinsic spin magnetic moment associated with the vector product $i{\bm d} \times {\bm d}^*$, 
where the complex vector ${\bm d}$ represents the triplet gap function. 
We show that a similar spin-polarized pairing appears in a helical metal proximitized by a current-biased conventional superconductor, 
as illustrated in Fig. \ref{f_j}a. Such a possibility is rather counter-intuitive since in conventional superconducting hybrids there is no pairing interaction in the triplet channel, 
and so ${\bm d}=0$. The roles of the ${\bm d}$ vector and the pair magnetic moment are assumed by 
the proximity-induced triplet pair amplitude, ${\bm f}$, and the product $i{\bm f} \times {\bm f}^*$ 
which describes the Cooper-pair spin polarization (CSP). 
We find that the CSP depends on the density, ${\bm j}$, of the applied supercurrent: 
$\overline{ 
i 
{\bm f}({\bm k},{\bm j}) \times {\bm f}^*({\bm k},{\bm j})
}\propto {\bm j} \times {\bm z}$, 
where the bar denotes averaging over the wave-vector (${\bm k}$) directions at the Fermi level, while ${\bm z}$ is the surface normal. 
The generation of the CSP is a form of the magnetoelectric effect. 
Unlike the Edelstein effect, which has a purely thermodynamic nature, 
the paradigm of the nonunitary pairing offers access to nonequilibrium magnetoelectric transport, allowing an electrical measurement of the CSP. 

To detect the CSP, we propose to measure the tunneling Andreev conductance, $G({\bm j})$, 
between the helical metal and a ferromagnet, as illustrated in Figs. \ref{f_j}c and d.
Such a junction acts as a spin valve in which the direction of ${\bm j}$ controls 
the orientation of the CSP with respect to the ferromagnetic magnetization ${\bm m}$. 
We show that reversing the supercurrent direction (${\bm j} \to -{\bm j}$) is equivalent to switching the magnetic configuration of the structure, 
which produces the change in the Andreev conductance proportional to the CSP:
\begin{equation}
G({\bm j}) - G(-{\bm j}) \propto 
(
\overline{ 
i 
{\bm f}({\bm k},{\bm j}) \times {\bm f}^*({\bm k},{\bm j})
}
)
\cdot 
{\bm m} 
\propto 
{\bm z} \cdot ( {\bm m} \times {\bm j}).
\nonumber
\end{equation}
The conductance difference emerges from the triplet Andreev reflection generated by the supercurrent, 
which can be viewed as the magnetoelectric Andreev effect.

{\em Electrically induced non-unitary triplet pairing.}
The generic Hamiltonian of a helical metal (HM) is 
$\hat{h}_{_{HM}}=\xi_{\bm k} + {\bm \sigma} \cdot {\bm \Omega}_{\bm k}$. 
It consists of a spin-independent part $\xi_{\bm k}=\epsilon_{\bm k} - \mu$ and the SOC where 
${\bm \sigma}$ is the Pauli matrix vector and ${\bm \Omega}_{\bm k}$ is a band-structure-dependent SO field; 
$\mu$ is the chemical potential and ${\bm k}=[k_x,k_y,0]$. 
We will do the calculations for arbitrary ${\bm \Omega}_{\bm k}$ and $\epsilon_{\bm k}$, 
illustrating the main points for the Dirac surface state with ${\bm \Omega}_{\bm k} = {\cal A} {\bm k} \times {\bm z}$ and $\epsilon_{\bm k}=0$, where ${\cal A}/\hbar$ is the carrier velocity.

In a contact with a conventional superconductor (see Fig. \ref{f_j}a), a HM acquires superconducting pair correlations 
which can be described by the Bogoliubov-de Gennes Hamiltonian: 
\begin{eqnarray}
\hat{H} =
\left[
\begin{array}{cc}
\xi_{{\bm k} + {\bm q}} + {\bm \sigma} \cdot {\bm \Omega}_{ {\bm k} + {\bm q} }  & \Delta i\sigma_y \\
-\Delta i\sigma_y & - (\xi_{-{\bm k} + {\bm q}} + {\bm \sigma} \cdot {\bm \Omega}_{-{\bm k} + {\bm q}})^*
\end{array}
\right],
\label{H_HM}
\end{eqnarray}
where $\Delta$ is the induced $s$-wave pair potential and the wave-vector shift ${\bm q}$ accounts for a uniform phase gradient  
$\nabla\varphi =2{\bm q}$. We assume that the latter is created by a homogeneous supercurrent density ${\bm j}= \nu_{_S} \nabla\varphi$ in the S,
so that ${\bm q}={\bm j}/(2\nu_{_S})$, where $\nu_{_S}$ is the condensate stiffness constant. For the Dirac surface state, 
Eq. (\ref{H_HM}) yields the energy spectrum $E_{\bm k}= {\cal A} {\bm n} \cdot {\bm q} \pm \sqrt{ ({\cal A}|{\bm k}|  - \mu)^2 + |\Delta|^2 }$ 
with ${\bm n}={\bm k}/|{\bm k}|$. We are interested in the small-gradient case ${\cal A}|{\bf q}| \ll \Delta$ 
when the HM is fully gapped at $E=0$.

To classify the induced superconducting correlations we consider the matrix pair amplitude:
\begin{equation}
\Big\langle
\begin{matrix}
a_{\uparrow {\bm k} }(t) a_{\uparrow -{\bm k} }(t) \,\,\,   
a_{\uparrow {\bm k} }(t) a_{\downarrow -{\bm k} }(t)   \\
a_{\downarrow {\bm k} }(t) a_{\uparrow -{\bm k} }(t)\,\,\,   
a_{\downarrow {\bm k} }(t) a_{\downarrow -{\bm k} }(t)
\end{matrix}
\Big\rangle
=
[f_0(t,{\bm k})+
{\bm \sigma}{\bm f}(t,{\bm k})]i\sigma_y.
\nonumber
\end{equation}
Here, $a_{s {\bm k}}(t)$ is the annihilation operator for an electron with spin projection $s=\uparrow,\downarrow$ and momentum ${\bm k}$ at time $t$; 
$\langle...\rangle$ denotes the ground-state expectation value. We use the singlet-triplet basis, with the singlet pair amplitude 
$f_0(t,{\bm k})$ and the triplet vector
${\bm f}(t,{\bm k}) =[ (f_{\downarrow\downarrow} - f_{\uparrow\uparrow})/2, -i(f_{\uparrow\uparrow} + f_{\downarrow\downarrow})/2, f_{\uparrow\downarrow +\downarrow\uparrow}]$ 
combining the amplitudes $f_{\uparrow\uparrow}(t,{\bm k})$, $f_{\downarrow\downarrow} (t,{\bm k})$, and $f_{\uparrow\downarrow +\downarrow\uparrow}(t,{\bm k})$
of the triplet states with total spin projections $S_z=1,-1$, and $0$ \cite{Annunziata12,Fritsch14}. 

We have calculated all pairing amplitudes from the condensate Green function for Eq. (\ref{H_HM}) \cite{SOM}. 
The result for the ${\bm f}$ vector is   
$
{\bm f}(t,{\bm k},{\bm q})=(i/2\pi)\int {\bm f}(E,{\bm k},{\bm q}) dE, 
$
where the integration goes over energy $E$ and involves a complex energy-dependent ${\bm f}$ vector
\begin{eqnarray}
&&
{\bm f}(E,{\bm k},{\bm q}) = \frac{\Delta}{\Pi(E,{\bm k},{\bm q})} 
[
E ({\bm \Omega}_{ {\bm k} + {\bm q} } + {\bm \Omega}_{-{\bm k} + {\bm q}}) +
\nonumber\\
&&
\xi_{-{\bm k} + {\bm q}}\, {\bm \Omega}_{ {\bm k} + {\bm q} } - \xi_{{\bm k} + {\bm q}}\, {\bm \Omega}_{-{\bm k} + {\bm q}} + 
i {\bm \Omega}_{ {\bm k} + {\bm q} }\times {\bm \Omega}_{-{\bm k} + {\bm q}}
],
\label{f_Ek_2}
\end{eqnarray}
where $\Pi(E,{\bm k},{\bm q})$ is a real function \cite{SOM}. Noteworthy is the imaginary term in Eq. (\ref{f_Ek_2}). 
It can be interpreted as the mutual {\em torque} of the particle and hole spins 
due to the misalignment of the SO fields ${\bm \Omega}_{ {\bm k} + {\bm q} }$ and ${\bm \Omega}_{ -{\bm k} + {\bm q} }$. 
This torque is generated by a supercurrent and 
is the reason for the non-unitary pairing in our model. 
If ${\bm \Omega}_{ \pm {\bm k} + {\bm q} }$ lie in the HM plane, the imaginary part of ${\bm f}$ points out of the plane, 
yielding the amplitude of the $S_z=0$ triplet (see also Fig. \ref{f_j}a).
The real part of ${\bm f}$ lies in the HM plane and describes the mixture of the equal-spin triplets.
For specific ${\bm \Omega}_{\bm k}$ and $\xi_{\bm k}$ Eq.(\ref{f_Ek_2}) 
recovers relevant earlier results on the proximity-induced paring 
in HMs \cite{Stanescu10,Potter11,Yokoyama12,BlackSchaffer12,Maier12,GT13,Snelder15,GT15,Triola16}.  

By analogy with non-unitary triplet superconductors \cite{Leggett75,Sigrist91}, 
the vector product $i{\bm f}(E,{\bm k},{\bm q}) \times {\bm f}^*(E,{\bm k},{\bm q})$ can be used to measure the CSP. 
We calculate it in the linear approximation with respect to ${\bm q}$ and average over the ${\bm k}$ directions at the Fermi level \cite{SOM}: 
\begin{eqnarray}
\frac{
\overline{i {\bm f}(0,{\bm k},{\bm q}) \times {\bm f}^*(0,{\bm k},{\bm q})} 
}
{4|\Delta|^2}
& \approx & 
\overline{
\frac{
\xi_{\bm k}
\,
{\bm \Omega}_{\bm k} \times [ {\bm \Omega}_{\bm k} \times ({\bm q}\cdot \partial_{\bm k}) {\bm \Omega}_{\bm k}]
}{\Pi(0,{\bm k},0)^2}
}
\nonumber\\
&=&
\frac{ {\cal A}^3 {\bm k}^2 \mu}{2\Pi(0,{\bm k},0)^2}\, ({\bm q} \times {\bm z}).
\label{ff*_Dirac}
\end{eqnarray}
The emergence of the CSP in response to the supercurrent indicates the magnetoelectric effect. 
For the Dirac surface state [see Eq. (\ref{ff*_Dirac})], the CSP direction is perpendicular 
to the applied supercurrent (see also Fig. \ref{f_j}b).
This resembles the current-induced thermodynamic magnetization of a Rashba SO-coupled S \cite{Edelstein95,Yip02,Bauer12,Konschelle15}.    
We argue that the CSP in Eq. (\ref{ff*_Dirac}) 
is a different observable which is related to the magnetoelectric Andreev transport.

\begin{figure}[t]
\includegraphics[width=45mm]{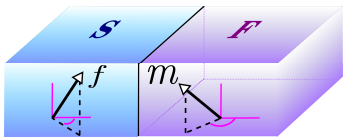}
\caption{(Color online) Schematic of a superconductor (S)/ferromagnet (F) tunnel junction in our model of Andreev reflection.
We consider non-collinear triplet vector ${\bm f}$ (in S) and magnetization vector ${\bm m}$ (in F).
}
\label{f_m}
\end{figure}

{\em How does CSP manifest itself in Andreev reflection?} 
To answer this question, we consider a tunnel junction between a ferromagnet (F) and a superconductor (S) with a generic condensate Green function
$
\hat{F}(E,{\bm k}) = [f_0(E,{\bm k}) + {\bm \sigma}{\bm f}(E,{\bm k})]i\sigma_y
$
(see also Fig. \ref{f_m}).
The junction is described by the tunneling Hamiltonian 
$
H_T =
\sum_{
s {\bm k} {\bm k^\prime} 
} 
\left[ b^\dagger_{s {\bm k}} \, t_{{\bm k},{\bm k^\prime} } \, a_{s {\bm k^\prime} }
+  
a^\dagger_{s {\bm k^\prime} } \, t^*_{ {\bm k},{\bm k^\prime} } \, b_{s {\bm k}} 
\right],
$
where $a^\dagger_{s {\bm k^\prime} }$ and $a_{s {\bm k^\prime}}$ are the creation and annihilation operators in the S,  
$b^\dagger_{s {\bm k} }$ and $b_{s {\bm k}}$ are the analogous operators in the F, and $t_{{\bm k},{\bm k^\prime} }$ 
is the spin-independent tunneling matrix element obeying time-reversal symmetry: $t^*_{ {\bm k},{\bm k^\prime} }=t_{ -{\bm k},-{\bm k^\prime} }$. 
A finite bias voltage $V$ at the junction generates the tunneling current  
$I= (ie/\hbar)
\sum_{
s {\bm k} {\bm k^\prime} 
} 
\left[ t_{{\bm k},{\bm k^\prime} } \langle b^\dagger_{s {\bm k}} a_{s {\bm k^\prime} }\rangle
-  
t^*_{ {\bm k},{\bm k^\prime} } \langle a^\dagger_{s {\bm k^\prime} } b_{s {\bm k}} \rangle  
\right]$,
where the brackets $\langle ...\rangle$ denote the non-equilibrium expectation values.   
To calculate $I$, we use perturbation theory with respect to $H_T$ combined with 
the non-equilibrium Green function approach of Ref. \cite{Cuevas96} and its version for conventional S/F junctions \cite{Tkachov02}. 
This method relates the current to the Green functions of the leads, enabling the treatment of the proximity structures such as the S/HM hybrids, 
where the triplet Andreev processes are associated with the proximity-induced pair amplitude ${\bm f}(E,{\bm k})$ rather than the ${\bm d}$ vector 
(cf., e.g., Refs. \cite{Honerkamp98,Kastening06,Linder07}). 
Assuming that the single-particle transport is suppressed by the excitation gap, we focus on the Andreev reflection (AR) contribution, $I_{_A}$, 
which appears in the fourth perturbation order and has the form (cf. \cite{Blonder82}):  
$I_{_A}=(2e/h)\int A(E)[n(E-eV)-n(E)]dE$, where $n(E)$ is the Fermi distribution function and 
$A(E)$ is the AR probability given by
%
%
\begin{eqnarray}
&
A(E) =
\pi^2 \sum_{{\bm k}_1{\bm k}_2{\bm k}{\bm k}^\prime} 
t_{{\bm k}_1,{\bm k}} t^*_{{\bm k}_2,{\bm k}} \left(t_{{\bm k}_1,{\bm k}^\prime} t^*_{{\bm k}_2,{\bm k}^\prime}\right)^*
\times
&
\nonumber\\
&
{\rm Tr}
\left[
\hat{\rho}_{_F}(E,{\bm k}_1) \hat{F}(E,{\bm k}) \hat{\rho}^*_{_F}(-E,-{\bm k}_2) \hat{F}^\dagger(E,{\bm k}^\prime) 
\right].
&
\label{A}
\end{eqnarray}
This equation illustrates the general AR mechanism, whereby a particle [represented by the spectral function $\hat{\rho}_{_F}(E,{\bm k})$] 
is converted into a Fermi-sea hole propagating back in time [hence, the spectral function $\hat{\rho}^*_{_F}(-E,-{\bm k})$], 
while a Cooper pair is created in the S. The latter is described by the condensate Green function $\hat{F}(E,{\bm k})$ and 
its hermitian conjugate $\hat{F}^\dagger(E,{\bm k})$ 
\footnote{Equation (\ref{A}) admits a straightforward generalization in terms of the spatially resolved Green functions \cite{Cuevas96}. 
This would not change our results qualitatively because they reflect the bulk properties of the S and F.}. 
The hat indicates $2\times 2$ spin matrices and ${\rm Tr}$ is the corresponding trace operation. 
The F is modeled by the Stoner Hamiltonian $\hat{h}_{_F} = \eta_{\bm k} - J {\bm m} \cdot {\bm \sigma}$, 
where the bare band dispersion $\eta_{\bm k}$ is split into the majority ($\eta_{\bm k} - J$) and minority ($\eta_{\bm k} + J$) spin bands 
by the exchange interaction parametrized by energy $J>0$ and the unit vector ${\bm m}$ specifying the magnetization direction.
The F spectral function can be obtained from the retarded Green function as
$
\hat{\rho}_{_F}(E,{\bm k})=\sum_{\alpha=\pm} \rho_\alpha (E,{\bm k}) \hat{P}_\alpha, 
$
where $\alpha= \pm$ denote the spin projections of the majority and minority states, respectively, on the magnetization ${\bm m}$;  
$\rho_\alpha(E,{\bm k})=\delta(E - \eta_{\bm k} + \alpha J)$ and $\hat{P}_\alpha = \frac{1}{2}(1 + \alpha {\bm \sigma}\cdot {\bm m})$ 
are the spectral density and the spin projector for the majority/minority states. 

After evaluating $A(E)$ in Eq. (\ref{A}), we use it to calculate the zero-bias and -temperature AR conductance.
Here we just quote the final result \cite{SOM}:   
\begin{widetext}
\begin{eqnarray}
G =\frac{2e^2}{h}A(0) &=& \frac{2e^2\pi^2}{h}\sum\limits_{{\bm k}_1 {\bm k}_2 \alpha=\pm }
\Bigl|
\sum_{ {\bm k} }t_{{\bm k}_1,{\bm k}}t^*_{{\bm k}_2,{\bm k}}
\left[
f_0(0,{\bm k}) + \alpha {\bm f}(0,{\bm k})\cdot {\bm m}
\right]
\Bigr|^2
\rho_\alpha(0,{\bm k}_1)\rho_{-\alpha}(0,-{\bm k}_2)
\nonumber\\
& &
\label{G}\\
&+&
\frac{e^2\pi^2}{h}\sum\limits_{{\bm k}_1 {\bm k}_2 \alpha=\pm }
\Bigl\|
\sum_{ {\bm k} }t_{{\bm k}_1,{\bm k}}t^*_{{\bm k}_2,{\bm k}}
\left[
{\bm m} \times ({\bm f}(0,{\bm k}) \times {\bm m})
- i\alpha 
{\bm f}(0,{\bm k}) \times {\bm m}
\right]
\Bigr\|^2
\rho_\alpha(0,{\bm k}_1)\rho_\alpha(0,-{\bm k}_2).
\nonumber
\end{eqnarray}
\end{widetext}
The two terms in Eq. (\ref{G}) correspond to two different types of AR. 
One is the opposite-spin AR in which the hole spin projection $-\alpha$ is antiparallel to that of the particle, $\alpha$ [see first term in Eq. (\ref{G})]. 
This process creates the Cooper pair in a mixed singlet-triplet state with the zero total spin projection on the magnetization direction ${\bm m}$.
The other process [see second term in Eq. (\ref{G})] involves the particle and the hole with equal spin projections $\alpha$. 
In this case, the Cooper pair is created in a triplet state with the total spin projection $1$ on the magnetization direction ${\bm m}$. 
Such equal-spin AR occurs for a non-collinear orientation of the ${\bm f}$ and ${\bm m}$ vectors and 
is accompanied by the transfer of the spin angular momentum $\hbar$ and torques on the magnetization.    
The symbol $\|...\|$ above means the norm of a complex vector, e.g., $\|{\bm V}\| = \sqrt{ {\bm V}\cdot {\bm V}^*}$. 
There is a close mathematical analogy between the ${\bm f}$ vector and the spin accumulation in normal metals, which exerts similar torques on the magnetization 
(see, e.g., review \cite{Brataas06} and Refs. \cite{Zhao08,Linder09,Wu14,Alidoust15,Holmqvist16}). 

The equal-spin AR in Eq. (\ref{G}) depends on the CSP. For concreteness,  
let us consider the AR process that results in the injection of an electron pair from the S into the F. 
Since the two electrons produce the torques independently, one of the torques can have the form 
${\bm m} \times ({\bm f}(0,{\bm k}) \times {\bm m})$, while the other ${\bm f}^*(0,{\bm k}) \times {\bm m}$. 
Here the complex conjugation reflects the time-reversal relation within the pair.    
The product of these torques is  
$[{\bm m} \times ({\bm f}(0,{\bm k}) \times {\bm m})]
\cdot 
[{\bm f}^*(0,{\bm k}) \times {\bm m}] =
{\bm m} \cdot [{\bm f}(0,{\bm k}) \times {\bm f}^*(0,{\bm k}]$,
which yields the CSP projected on the magnetization direction. 
Using the notation $G_{odd}$, we can write the CSP-dependent part of Eq. (\ref{G}) as \cite{SOM}:
\begin{eqnarray}
G_{odd} = \frac{2e^2}{h}
\,
{\bm m} \cdot 
\sum_{ {\bm k}{\bm k}^\prime\alpha=\pm}
\alpha 
\left|
\Gamma^\alpha_{ {\bm k},{\bm k}^\prime}
\right|^2
i{\bm f}(0,{\bm k}) \times {\bm f}^*(0,{\bm k}^\prime),
\,\,\,
\label{G_odd}
\end{eqnarray}
where we introduce the energy scale
$\Gamma^\alpha_{ {\bm k}, {\bm k}^\prime} = \pi \sum_{{\bm k}_1} t_{{\bm k}_1,{\bm k}}t^*_{{\bm k}_1,{\bm k}^\prime}\rho_\alpha(0,{\bm k}_1)$ 
characterizing the normal-state tunneling rates into the majority and minority F states. 
The wave-vectors of the tunneling electrons are generally different, as seen from Eq. (\ref{G_odd}). 
This result takes a simpler form for random tunneling 
which suppresses the incoherent terms with ${\bm k}^\prime \not = {\bm k}$ \cite{SOM}:
\begin{eqnarray}
G_{odd} =
\frac{2e^2}{h}
\Bigl(
\Gamma^2_+
- 
\Gamma^2_-
\Bigr) 
\,
{\bm m} \cdot 
\sum_{\bm k}
i{\bm f}(0,{\bm k}) \times {\bm f}^*(0,{\bm k}),
\label{G_odd_av}
\end{eqnarray}
where $\Gamma_\pm$ are the ${\bm k}$-independent tunneling rates \cite{SOM}. 
The vector $i{\bm f}(0,{\bm k}) \times {\bm f}^*(0,{\bm k})$ couples directly to the magnetization ${\bm m}$ and is, therefore, 
a valid observable characterizing the CSP (see also Ref. \cite{Hillier12}). 

\begin{figure}[t]
\includegraphics[width=85mm]{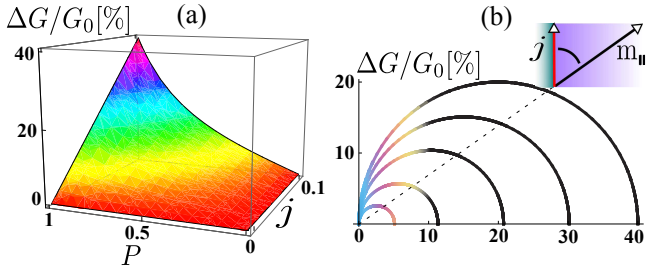}
\caption{(Color online)
(a) Magnetoelectric Andreev conductance ratio $\Delta G/G(0) = [G({\bm j}) - G(-{\bm j})]/G(0)$ versus supercurrent density 
$j$ and spin polarization of the ferromagnet, $P$ [see also Eq. (\ref{dG_P})]. 
$j$ is measured in units of $2\nu_{_S}\mu/{\cal A}$. 
(b) Polar plot of the conductance ratio as function of the angle between 
the in-plane magnetization ${\bm m}_\|$ and supercurrent ${\bm j}$ 
for different values of the spin polarization $P$.
}
\label{dG_fig}
\end{figure}

{\em Magnetoelectric Andreev effect.}
Let us apply the above results to the S/HM/F hybrids shown in Figs. \ref{f_j}c and d.
From Eqs. (\ref{ff*_Dirac}) and (\ref{G_odd_av}), it is clear that $G_{odd}$   
is an odd function of the supercurrent ${\bm j}=2\nu_{_S}{\bm q}$. 
This makes the total conductance $G({\bm j})$ asymmetric with respect to the direction of ${\bm j}$. 
The difference between $G({\bm j})$ and $G(-{\bm j})$ is simply twice the odd conductance (\ref{G_odd_av}), i.e.  
$G({\bm j}) - G(-{\bm j})=2G_{odd}({\bm j})$ \cite{SOM}. 
This shows that the spin-polarized AR conductance is generated by an electric supercurrent in an amount proportional to the CSP,   
which we call here the magnetoelectric Andreev effect (MAE).   
To estimate the MAE we calculate the ratio of $G({\bm j}) - G(-{\bm j})$ to the zero-current conductance $G(0)$ \cite{SOM}:
\begin{eqnarray}
&&
\frac{G({\bm j})-G(-{\bm j})}{G(0)} = {\cal A} \frac{{\bm z} \cdot ( {\bm m} \times {\bm j} )}{2\nu_{_S}\mu} 
\nonumber\\
&&
\times
\frac{4P}{1 + P^2{\bm m}^2_\perp + (1-P^2){\cal F}_s\left(\frac{\Delta}{\mu}\right)/{\cal F}_t \left( \frac{\Delta}{\mu}\right)}, 
\label{dG_P}
\end{eqnarray}
where $P = \frac{ \rho_+ - \rho_- }{\rho_+ + \rho_-}$ is the F spin polarization 
given by the densities of the majority, $\rho_+$, and minority, $\rho_- $, states at the Fermi level, 
${\bm m}_\perp$ is the out-of-plane unit vector magnetization component, and 
the functions ${\cal F}_{s,t}(x)= 1 +\frac{1 \pm x^2}{2x}\left( \frac{\pi}{2} + \arctan\frac{1-x^2}{2x} \right)$ 
come from the ${\bm k}$ integrals in the singlet ($s$) and triplet ($t$) terms 
of Eq. (\ref{G}).  
  
As shown in Fig. \ref{dG_fig}a, the ratio (\ref{dG_P}) attains sizable values 
with increasing supercurrent $j$ and spin polarization $P$. 
The highest value of 40$\%$ is achieved for a fully spin polarized F ($P=1$) 
at a reasonable current density corresponding to a small fraction of the Fermi energy: 
${\cal A} j/(2\nu_{_S})=0.1 \mu$. The MAE is anisotropic, 
depending on the orientation of the F magnetization with respect to the supercurrent, and 
is largest when ${\bm m}$ is perpendicular to ${\bm j}$.  
We illustrate this in Fig. \ref{dG_fig}b by plotting the ratio (\ref{dG_P}) as function 
of the angle between the in-plane magnetization ${\bm m}_\|$ and ${\bm j}$.   
Another source of the anisotropy is the dependence on the out-of-plane component 
${\bm m}_\perp$ in Eq. (\ref{dG_P}). 
This dependence is not related to the CSP, coming from the zero-current conductance $G(0)$ \cite{SOM}  
in agreement with the results of Ref. \cite{Hoegl15} for a Rashba SO-coupled S/F interface. 

In conclusion, we have studied the magnetoelectric effect in Andreev transport    
in superconducting hybrids based on two-dimensional helical metals. 
The effect stems from the lack of the inversion symmetry and is expected in a rather wide class of non-centrosymmetric materials.   
Due to a strong spin-orbit coupling in topological insulators, their Dirac-like surface states can be particularly
suitable for an experimental realization of our proposal.
The magnetoelectric Andreev effect is intimately related to the non-unitary triplet pairing. 
The latter plays an important role in understanding unconventional pairing mechanisms (see, e.g., Ref. \cite{Hillier12}), 
but has not been observed in transport yet. Our findings can lead to significant advances 
in achieving electrically controlled superconducting spintronics and diagnostics of unconventional superconducting states.

\acknowledgments
The author thanks F. S. Bergeret, I. V. Tokatly, J. F. Annett, and A. V. Balatsky for discussions 
and the German Research Foundation (DFG) for financial support (Grants No TK60/1-1 and TK60/4-1).

\newpage

\onecolumngrid

\begin{center}
{\large\bf Supplemental material to}
\vskip 0.5cm

{\Large\bf "Magnetoelectric Andreev effect due to proximity-induced non-unitary triplet superconductivity in helical metals"}

\vskip 0.25cm

by G. Tkachov

\vskip 0.25cm

\textit{
Institute for Theoretical Physics and Astrophysics, Wuerzburg University, Am Hubland, 97074 Wuerzburg, Germany
}
\end{center}

\section{\large A. Induced non-unitary triplet superconductivity. Theoretical framework and Properties}

In this section, we analyze in detail the specific type of spin-triplet Cooper pairing arising from the Edelstein effect (EE). 
Our goal is to point out a surprising connection between this EE-related pairing and the paradigmatic non-unitary pairing (NUP) 
in intrinsic triplet superconductors with a complex ${\bm d}$ vector satisfying $i{\bm d}\times {\bm d}^*\not=0$ \cite{Leggett75,Sigrist91}. 
To that end, we generalize the model used in the main text by including a complex ${\bm d}$ vector into the Bogoliubov-de Gennes (BdG) Hamiltonian. 
This allows a unified description of the EE-related pairing and the intrinsic NUP in terms of the anomalous (Gorkov) Green function $\hat{F}(E,{\bm k})$ of the BdG equation.
The use of $\hat{F}(E,{\bm k})$ is also motivated by two other circumstances. 
First, $\hat{F}(E,{\bm k})$ is directly related to observable transport properties, and, second,  
it describes also the proximity-induced superconductivity in non-interacting systems which do not possess the superconducting gap function per se as, e.g., 
in the case considered in the main text.
We identify the key common feature of the EE-related pairing and the intrinsic NUP: 
they are both characterized by the nonvanishing axial vector $i{\bm f}(E,{\bm k}) \times {\bm f}^*(E,{\bm k})$ related to the condensate spin polarization 
[where ${\bm f}(E,{\bm k})$ is the vector triplet pair amplitude]. The use of the ${\bm f}$ vector allows one to extend the paradigm of the NUP beyond its original context, viz., to introduce 
the notion of the {\em induced NUP}. The latter can be defined as the {\em triplet pairing possessing the spin polarization $i{\bm f}(E,{\bm k}) \times {\bm f}^*(E,{\bm k})$ 
irrespective of the intrinsic ${\bm d}$ vector}. The EE-related pairing falls precisely into the latter category because 
it is induced from the singlet pairing through the combined effect of the spin-orbit coupling (SOC) and supercurrent, i.e. even for ${\bm d}\not=0$. 
This analysis substantiates two essential points:
\begin{itemize}
\item 
The NUP can exist as an {\em induced} state, and the singlet superconductor (S)/helical metal (HM) hybrids, where ${\bm d}=0$, provide a minimal model to study this 
previously unexplored state.  
 
\item
The induced NUP is also expected in non-centrosymmetric superconductors with a mixed singlet-triplet pair potential (i.e. for ${\bm d}\not=0$), 
although there additional effects due to the non-collinearity of the ${\bm d}$ and SOC vectors come into play.  
\end{itemize}

\subsection{\large A1. Anomalous Green function for generic SOC and ${\bm d}$ vectors}

Compared to the main text, here we adopt a more general BdG Hamiltonian which has a mixed singlet-triplet pair potential $\hat{\Delta}$:
\begin{eqnarray}
\hat{H} =
\left[
\begin{array}{cc}
\xi_{{\bm k} + {\bm q}} + {\bm \sigma} \cdot {\bm \Omega}_{ {\bm k} + {\bm q} }  & \hat{\Delta} \\
\hat{\Delta}^\dagger  & - (\xi_{-{\bm k} + {\bm q}} + {\bm \sigma} \cdot {\bm \Omega}_{-{\bm k} + {\bm q}})^*
\end{array}
\right],
\qquad 
\hat{\Delta} = (\Delta + {\bm d} \cdot {\bm \sigma} )i\sigma_y.
\label{H}
\end{eqnarray}
Here, $\hat{\Delta}$ includes a (generally complex) ${\bm d}$ vector which, for intrinsic superconductors, must be determined self-consistently 
from the gap equation (see, e.g., Ref. \cite{Bauer12}). 
We leave this issue aside, focusing rather on the generic structure of the pair correlations described by the Green function, $\hat{\cal G}(E,{\bm k})$, of the BdG equation:
\begin{equation}
(E  - \hat{H}) \hat{\cal G}(E,{\bm k}) =  \hat{I},
\label{Eq_G}
\end{equation}
where $\hat{I}$ is the unit matrix in both particle-hole and spin sectors. We adopt the standard notations for the matrix elements of $\hat{\cal G}(E,{\bm k})$
in the particle-hole space: 
\begin{equation}
\hat{\cal G}(E,{\bm k})=
\left[
\begin{array}{cc}
\hat{G}(E,{\bm k})  &  \hat{F}(E,{\bm k}) \\
\hat{F}^\dagger(E,{\bm k}) &  \hat{\overline{G}}(E,{\bm k})
\end{array}
\right],
\label{G_ph}
\end{equation}
where each entry is a $2\times 2$ matrix in spin space: 
$\hat{G}(E,{\bm k})$ and $\hat{\overline{G}}(E,{\bm k})$ are the quasiparticle Green functions in the particle and hole sectors, respectively, 
$\hat{F}(E,{\bm k})$ and its hermitian conjugate $\hat{F}^\dagger(E,{\bm k})$ are the anomalous (Gorkov) Green functions. 
In the particle-hole space, Eq. (\ref{Eq_G}) can be written as

\begin{equation}
\left[
\begin{array}{cc}
E_+ - {\bm \sigma} \cdot {\bm \Omega}_+  & - (\Delta + {\bm d} \cdot {\bm \sigma} )i\sigma_y \\
i\sigma_y (\Delta^* + {\bm d}^* \cdot {\bm \sigma} ) & E_-  + {\bm \sigma}^* \cdot {\bm \Omega}_-
\end{array}
\right]
\left[
\begin{array}{cc}
\hat{G}  &  \hat{F} \\
\hat{F}^\dagger &  \hat{\overline{G}}
\end{array}
\right]
= 
\left[
\begin{array}{cc}
\hat{1}  &  0 \\
0 &  \hat{1}
\end{array}
\right],
\label{Eq_G_1}
\end{equation}
where we introduce the shorthand notations 
\begin{equation}
E_+ \equiv E -\xi_{{\bm k} + {\bm q}}, \qquad E_- \equiv E + \xi_{-{\bm k} + {\bm q}}, \qquad {\bm \Omega}_\pm \equiv {\bm \Omega}_{ \pm {\bm k} + {\bm q} },
\end{equation}
and $\hat{1}$ for the unit matrix in spin space. 

Our goal is to calculate the function $\hat{F}$ describing all pairing types arising from the spin, orbital and energy degrees of freedom.  
To that end, we consider the pair of equations for $\hat{F}$ and $\hat{\overline{G}}$:
\begin{eqnarray}
&&
(E_+ - {\bm \sigma} \cdot {\bm \Omega}_+) \hat{F} - (\Delta + {\bm d} \cdot {\bm \sigma} )i\sigma_y \hat{\overline{G}} = 0,
\label{Eq_12}\\
&&
i\sigma_y (\Delta^* + {\bm d}^* \cdot {\bm \sigma} ) \hat{F} + (E_-  + {\bm \sigma}^* \cdot {\bm \Omega}_-) \hat{\overline{G}} = \hat{1}.
\label{Eq_22}
\end{eqnarray}
It is convenient to introduce the function $\hat{G}^\prime = i\sigma_y \hat{\overline{G}}$ and multiply Eq. (\ref{Eq_22}) by $i\sigma_y$ from left:

\begin{eqnarray}
&&
(E_+ - {\bm \sigma} \cdot {\bm \Omega}_+) \hat{F} - (\Delta + {\bm d} \cdot {\bm \sigma} ) \hat{G}^\prime = 0,
\label{Eq_12_1}\\
&&
-(\Delta^* + {\bm d}^* \cdot {\bm \sigma} ) \hat{F} + (E_-  - {\bm \sigma} \cdot {\bm \Omega}_-) \hat{G}^\prime = i\sigma_y.
\label{Eq_22_1}
\end{eqnarray}
We can now multiply Eq. (\ref{Eq_12_1}) by $(E_-  - {\bm \sigma} \cdot {\bm \Omega}_-)(\Delta - {\bm d} \cdot {\bm \sigma} )$ from left and exclude $\hat{G}^\prime$ with the help of Eq. (\ref{Eq_22_1}). 
In this way, we obtain a closed-form equation for $\hat{F}$: 

\begin{equation}
\left[ 
(E_-  - {\bm \sigma} \cdot {\bm \Omega}_-)(\Delta - {\bm d} \cdot {\bm \sigma} )(E_+ - {\bm \sigma} \cdot {\bm \Omega}_+) 
- (\Delta^2 -  {\bm d}^2)(\Delta^* + {\bm d}^* \cdot {\bm \sigma} ) 
\right]
\hat{F} =  (\Delta^2 -  {\bm d}^2)i\sigma_y.
 \label{Eq_F}
\end{equation}
After evaluating the products of the spin matrices, we can write Eq. (\ref{Eq_F}) as 
\begin{equation}
\left[ D_0 - {\bm D} \cdot {\bm \sigma} \right] \hat{F} =  (\Delta^2 -  {\bm d}^2)i\sigma_y,
 \label{Eq_F_1}
\end{equation}
where the scalar $D_0$ and vector ${\bm D}$ are given by

\begin{eqnarray}
D_0 &=& \Delta (E_+E_- + {\bm \Omega}_+ \cdot {\bm \Omega}_- - |\Delta|^2) + E_+ {\bm \Omega}_- \cdot {\bm d} + E_- {\bm \Omega}_+ \cdot {\bm d} 
+ i \left({\bm \Omega}_- \times \cdot {\bm \Omega}_+\right) \cdot {\bm d} + \Delta^* {\bm d}^2,
\label{D_0}\\
& &
\nonumber\\
{\bm D} &=& (\Delta E_+ + {\bm \Omega}_+ \cdot {\bm d}){\bm \Omega}_ - + (\Delta E_- + {\bm \Omega}_- \cdot {\bm d}){\bm \Omega}_+ + \Delta\, i {\bm \Omega}_+ \times {\bm \Omega}_-
\nonumber\\
& &
\nonumber\\
&+& 
\Delta^2 {\bm d}^*  + iE_+ {\bm d} \times {\bm \Omega}_ -  - iE_- {\bm d} \times {\bm \Omega}_ +  
+ (E_+ E_- - {\bm \Omega}_+ \cdot {\bm \Omega}_- - {\bm d} \cdot {\bm d}^*){\bm d} + {\bm d} \times (  {\bm d} \times  {\bm d}^*).
\label{D}
\end{eqnarray}
Inverting Eq. (\ref{Eq_F_1}) yields the condensate function $\hat{F}$ explicitly: 
\begin{flushleft}
\fbox{
\begin{minipage}{\textwidth}
\begin{equation}
\hat{F} =  (\Delta^2 -  {\bm d}^2)  \frac{D_0 + {\bm D} \cdot {\bm \sigma}}{D^2_0 - {\bm D}^2} i\sigma_y \equiv (f_0 + {\bm f} \cdot {\bm \sigma}) i\sigma_y,
 \label{F}
\end{equation}
where the singlet ($f_0$) and triplet (${\bm f}$) amplitudes are related to the scalar $D_0$ and vector ${\bm D}$, respectively, by
\begin{equation}
f_0 = \frac{\Delta^2 - {\bm d}^2}{D^2_0 - {\bm D}^2} 
\Bigl[ 
\Delta (E_+E_- + {\bm \Omega}_+ \cdot {\bm \Omega}_- - |\Delta|^2) + E_+ {\bm \Omega}_- \cdot {\bm d} + E_- {\bm \Omega}_+ \cdot {\bm d} 
+ i \left({\bm \Omega}_- \times {\bm \Omega}_+\right) \cdot {\bm d} + \Delta^* {\bm d}^2
\Bigr],
\label{f_0}
\end{equation}
\begin{eqnarray}
{\bm f} &=& \frac{\Delta^2 - {\bm d}^2}{D^2_0 - {\bm D}^2} 
\Bigl[
(\Delta E_+ + {\bm \Omega}_+ \cdot {\bm d}){\bm \Omega}_ - + (\Delta E_- + {\bm \Omega}_- \cdot {\bm d}){\bm \Omega}_+ + 
\overbrace{i \Delta\, {\bm \Omega}_+ \times {\bm \Omega}_-}^{\bf due \,\, to \,\, EE }
\nonumber\\
& &
\nonumber\\
&+& 
\Delta^2 {\bm d}^*  + iE_+ {\bm d} \times {\bm \Omega}_ -  - iE_- {\bm d} \times {\bm \Omega}_ +  
+ (E_+ E_- - {\bm \Omega}_+ \cdot {\bm \Omega}_- - {\bm d} \cdot {\bm d}^*){\bm d} +  {\bm d} \times 
\underbrace{( {\bm d} \times  {\bm d}^*)}_{\bf intrinsic \,\, NUP }
\Bigr].
\label{f}
\end{eqnarray}
\end{minipage}
}
\end{flushleft}

The last term in the second line of Eq. (\ref{f}) is characteristic of the intrinsic NUP.
It should be compared with the imaginary term in the first line of Eq. (\ref{f}) which comes from the EE.
The EE manifests itself as the mutual {\em torque} of the particle and hole spins 
coupled by the singlet pair potential as also seen from Eq. (\ref{Eq_F}) containing the product   
\begin{equation} 
\Delta ({\bm \sigma} \cdot {\bm \Omega}_+)({\bm \sigma} \cdot {\bm \Omega}_-) \equiv \Delta ({\bm \sigma} \cdot {\bm \Omega}_{ {\bm k} + {\bm q} })({\bm \sigma} \cdot {\bm \Omega}_{ -{\bm k} + {\bm q} })= 
\Delta {\bm \Omega}_{ {\bm k} + {\bm q} } \cdot  {\bm \Omega}_{ -{\bm k} + {\bm q} } + 
i \Delta {\bm \sigma} \cdot ( {\bm \Omega}_{ {\bm k} + {\bm q} } \times {\bm \Omega}_{ -{\bm k} + {\bm q} } ).
\label{coupling}
\end{equation}
The last (torque) term is generated by the supercurrent with ${\bm q}\not=0$, otherwise the SOC vectors are collinear by time-reversal symmetry, ${\bm \Omega}_{-{\bm k}} = - {\bm \Omega}_{\bm k}$.
We show next that both the intrinsic NUP and the EE produce the Cooper-pair spin polarization (CSP) in the form of the axial vector 
$i{\bm f}(E,{\bm k}) \times {\bm f}^*(E,{\bm k})$, allowing a unified description of these two effects.

\subsection{\large A2. Intrinsic NUP}

The intrinsic NUP as defined in Refs. \cite{Leggett75,Sigrist91} is recovered by setting  
$\Delta=0, \,\, {\bm \Omega}_\pm =0$, and ${\bm q}=0$ in all the equations above. 
This yields 
$
D_0=0, \,\, {\bm D}^2 = {\bm d}^2 \left[ (E_+ E_- - {\bm d} \cdot {\bm d}^*)^2 - (i{\bm d} \times  {\bm d}^*)^2 \right],
$
and, hence, the ${\bm f}$ vector (cf. Ref. \cite{Sigrist91})

\begin{equation}
{\bm f}(E,{\bm k}) = 
\frac{ (E^2 - \xi^2_{\bm k} - {\bm d} \cdot {\bm d}^*){\bm d} + {\bm d} \times ({\bm d} \times  {\bm d}^*)}
{ (E^2 - \xi^2_{\bm k} - \Delta^2_+)( E^2 - \xi^2_{\bm k} - \Delta^2_-)}, 
\qquad 
\Delta^2_\pm = {\bm d} \cdot {\bm d}^* \pm |i{\bm d} \times  {\bm d}^*|.
\label{f_intr_NUP}
\end{equation}
Here, $\Delta_\pm$ are the energy gaps for the pairing with the net spin projection along the vector $i{\bm d} \times  {\bm d}^*$ and opposite to it (the signs "$+$" and "$-$", respectively).
The paradigmatic example of the intrinsic NUP is the equal-spin pairing \cite{Leggett75} described by 
\begin{equation}
{\bm d}=\Delta_0({\bm k})\frac{{\bm x} +i{\bm y}}{2}, \qquad i{\bm d} \times {\bm d}^*=\frac{|\Delta_0({\bm k})|^2}{2} \, {\bm z}, \qquad \Delta_+ = |\Delta_0({\bm k})|, \qquad \Delta_-=0.
\label{ESP}
\end{equation}
In this example, the pairing occurs only between the spin-$\uparrow$ electrons, generating an excitation gap $|\Delta_0({\bm k})|$, 
while the spin-$\downarrow$ electrons remain unpaired and gapless. Above, ${\bm x}$, ${\bm y}$, and ${\bm z}$ are the cartesian unit vectors.
From Eq. (\ref{f_intr_NUP}) we can readily obtain the axial vector $i{\bm f}(E,{\bm k}) \times {\bm f}^*(E,{\bm k})$ as

\begin{equation}
i{\bm f}(E,{\bm k}) \times {\bm f}^*(E,{\bm k}) = 
\frac{ (E^2 - \xi^2_{\bm k})^2 - \Delta^2_+\Delta^2_-}
{ (E^2 - \xi^2_{\bm k} - \Delta^2_+)^2( E^2 - \xi^2_{\bm k} - \Delta^2_-)^2}
\, i{\bm d} \times {\bm d}^*. 
\label{ff*_intr_NUP}
\end{equation}
That is, $i{\bm d} \times {\bm d}^*\not=0$ generates the nonvanishing vector $i{\bm f}(E,{\bm k}) \times {\bm f}^*(E,{\bm k})$. 
The latter can be used on par with $i{\bm d} \times {\bm d}^*$ to describe the CSP.
While this is not surprising for the intrinsic NUP, the question arises whether the spin polarization $i{\bm f}(E,{\bm k}) \times {\bm f}^*(E,{\bm k})$ can exist 
irrespective of the ${\bm d}$ vector, e.g., in systems for which ${\bm d}=0$. Such a possibility is rather counter-intuitive and, as discussed next, 
is associated with a qualitatively different physics.
 
\subsection{\large A3. Induced NUP. Cooper-pair spin polarization due to Edelstein effect}

In the following, we consider the induced superconductivity in a non-interacting HM proximitized by a singlet S. 
This is an example of the system with a vanishing ${\bm d}$ vector, reflecting no intrinsic electron interactions in the triplet channel. 
In this case, the triplet correlations are induced from the singlet pairing through the SOC in the HM. Setting ${\bm d}=0$ in Eqs. (\ref{D_0}) -- (\ref{f}), 
we obtain the pairing amplitudes explicitly as  

\begin{eqnarray}
f_0(E, {\bm k}, {\bm q}) &=& \Delta
\frac{
(E - \xi_{{\bm k} + {\bm q}})(E + \xi_{-{\bm k} + {\bm q}}) + {\bm \Omega}_{ {\bm k} + {\bm q} } \cdot {\bm \Omega}_{ -{\bm k} + {\bm q} } -|\Delta|^2 
}
{
\Pi(E, {\bm k}, {\bm q})
},
\label{f_s}\\
& &
\nonumber\\
{\bm f}(E, {\bm k}, {\bm q}) &=& \Delta
\frac{
E ( {\bm \Omega}_{ {\bm k} + {\bm q} } + {\bm \Omega}_{ -{\bm k} + {\bm q} } ) + \xi_{-{\bm k} + {\bm q}} \, {\bm \Omega}_{ {\bm k} + {\bm q} } - \xi_{{\bm k} + {\bm q}} \, {\bm \Omega}_{ -{\bm k} + {\bm q} }
+ 
i {\bm \Omega}_{ {\bm k} + {\bm q} } \times {\bm \Omega}_{ -{\bm k} + {\bm q} }
}
{
\Pi(E, {\bm k}, {\bm q})
}.
\label{f_t}
\end{eqnarray}
The denominator is given by $\Pi(E, {\bm k}, {\bm q})= [ D^2_0(E, {\bm k}, {\bm q}) - {\bm D}^2(E, {\bm k}, {\bm q}) ]/\Delta^2$, which after some algebra reads 

\begin{eqnarray}
\Pi(E, {\bm k}, {\bm q}) &=& 
\left[(E-\xi_{{\bm k} + {\bm q}})^2 - {\bm \Omega}^2_{ {\bm k} + {\bm q} } -|\Delta|^2\right]
\left[(E+\xi_{-{\bm k} + {\bm q}})^2 - {\bm \Omega}^2_{ -{\bm k} + {\bm q} } -|\Delta|^2\right]
\nonumber\\
&+&
|\Delta|^2 \left[(\xi_{{\bm k} + {\bm q}} + \xi_{-{\bm k} + {\bm q}})^2 - ({\bm \Omega}_{ {\bm k} + {\bm q} } + {\bm \Omega}_{ -{\bm k} + {\bm q} } )^2 \right].
\label{Pi}
\end{eqnarray}
The zeros of this function yield the spectrum of the single-particle excitations in the HM. 
In the time-reversal symmetric case, where ${\bm q}=0$, ${\bm \Omega}_{ -{\bm k}} = -{\bm \Omega}_{ {\bm k}}$ and $\xi_{-{\bm k}}=\xi_{\bm k}$, the spectrum has an isotropic singlet gap $|\Delta|$.
It can be easily checked by factorizing $\Pi(E, {\bm k}, 0)$ as follows

\begin{eqnarray}
\Pi(E, {\bm k}, 0) &=& 
\left[(E-\xi_{\bm k})^2 - {\bm \Omega}^2_{\bm k} -|\Delta|^2\right]
\left[(E+\xi_{\bm k})^2 - {\bm \Omega}^2_{\bm k} -|\Delta|^2\right] +
4|\Delta|^2 \xi^2_{\bm k} 
\nonumber\\
&=& 
\left[
E^2 - (\xi_{\bm k} + |{\bm \Omega}_{\bm k}|)^2 - |\Delta|^2
\right]
\left[
E^2 - (\xi_{\bm k} - |{\bm \Omega}_{\bm k}|)^2 - |\Delta|^2
\right].
\label{Pi_TRS}
\end{eqnarray}
The latter equation yields two spin-split BCS-like spectral branches $E^{(1)}_{\bm k}=\pm \sqrt{ (\xi_{\bm k} + |{\bm \Omega}_{\bm k}|)^2 + |\Delta|^2 }$ and 
$E^{(2)}_{\bm k}=\pm \sqrt{ (\xi_{\bm k} - |{\bm \Omega}_{\bm k}|)^2 + |\Delta|^2 }$.

The results obtained here [cf. Eq. (2) of the main text] hold for generic spin-splitting field ${\bm \Omega}_{\bm k}$ and spin-independent dispersion $\xi_{\bm k}$.  
For specific ${\bm \Omega}_{\bm k}$ and $\xi_{\bm k}$ and in the appropriate limits, we recover the results for the Green functions obtained in 
Refs. \cite{Stanescu10,Potter11,Yokoyama12,BlackSchaffer12,Maier12,GT13,Snelder15,GT15,Triola16}.   
In the following, we focus on the CSP $i{\bm f}(E, {\bm k}, {\bm q}) \times {\bm f}^*(E, {\bm k}, {\bm q})$ 
which has not been studied before. Microscopically, it results from the particle-hole spin torque associated with the EE, 
i.e. is an induced effect as opposed to the intrinsic NUP [cf. Eq. (\ref{ff*_intr_NUP})].   
From Eq. (\ref{f_t}), we find the CSP at the Fermi level ($E=0$) as 
\begin{equation}
i{\bm f}(0, {\bm k}, {\bm q}) \times {\bm f}^*(0, {\bm k}, {\bm q}) = 
\frac{
2|\Delta|^2
}
{
\Pi(0,{\bm k},{\bm q})^2
}
(
\xi_{-{\bm k} + {\bm q}} \, {\bm \Omega}_{ {\bm k} + {\bm q} } - \xi_{{\bm k} + {\bm q}} \, {\bm \Omega}_{ -{\bm k} + {\bm q} }
)
\times 
(
{\bm \Omega}_{ {\bm k} + {\bm q} } \times {\bm \Omega}_{ -{\bm k} + {\bm q} }
).
\label{ff*_gen}
\end{equation}
The CSP can be generated solely by a supercurrent in the absence of any normal-state spin polarization
when ${\bm \Omega}_{\bm k}$ is a pure spin-orbit field restricted by time-reversal symmetry (${\bm \Omega}_{-{\bm k}}=-{\bm \Omega}_{\bm k}$).

The CSP (\ref{ff*_gen}) is odd in ${\bm q}$. We can therefore linearize it with respect to ${\bm q}$. 
Expanding ${\bm \Omega}_{ \pm {\bm k} + {\bm q} }\approx \pm {\bm \Omega}_{\bm k} + ({\bm q}\cdot \partial_{\bm k}){\bm \Omega}_{\bm k}$, 
we approximate the particle-hole torque by the lowest order term: 
${\bm \Omega}_{ {\bm k} + {\bm q} } \times {\bm \Omega}_{ -{\bm k} + {\bm q} } \approx 2 {\bm \Omega}_{\bm k} \times ({\bm q}\cdot \partial_{\bm k}){\bm \Omega}_{\bm k}$. 
In this order, the other factors in Eq. (\ref{ff*_gen}) should be taken at ${\bm q}=0$, which yields 

\begin{equation}
i {\bm f}(0,{\bm k},{\bm q}) \times {\bm f}^*(0,{\bm k},{\bm q}) \approx
\frac{4|\Delta|^2}{\Pi(0,{\bm k},0)^2}
\xi_{\bm k}
\,
{\bm \Omega}_{\bm k} \times [ {\bm \Omega}_{\bm k} \times ({\bm q}\cdot \partial_{\bm k}) {\bm \Omega}_{\bm k}].
\label{ff*_lin}
\end{equation}
Importantly, the CSP does not vanish upon averaging over the momentum directions. 
We prove this for the linear Dirac spectrum with ${\bm \Omega}_{\bm k} = {\cal A} {\bm k} \times {\bm z}$ and $\epsilon_{\bm k}=0$.
In this case, $({\bm q}\cdot \partial_{\bm k}) {\bm \Omega}_{\bm k} = {\cal A} ({\bm q} \times {\bm z})$, $\xi_{\bm k}=-\mu$ and, therefore,

\begin{equation}
i {\bm f}(0,{\bm k},{\bm q}) \times {\bm f}^*(0,{\bm k},{\bm q}) \approx
-\frac{4|\Delta|^2 {\cal A}^3\mu}{\Pi(0,k,0)^2}
({\bm k} \times {\bm z})
\times 
[({\bm k} \times {\bm z}) \times ({\bm q} \times {\bm z})] = 
-\frac{4|\Delta|^2 {\cal A}^3\mu}{\Pi(0,k,0)^2}
{\bm k} \, 
[{\bm k} \cdot ( {\bm z} \times {\bm q} )].
\label{ff*_lin_Dirac}
\end{equation}
The denominator depends only on the wave-vector magnitude $k=|{\bm k}|$ [see Eq. (\ref{Pi_TRS})],

\begin{eqnarray}
\Pi(0, k, 0) = \left(\mu^2 -|\Delta|^2 - {\cal A}^2 k^2 \right)^2 + 4|\Delta|^2\mu^2,
\label{Pi_Dirac}
\end{eqnarray}
so the averaging affects only the numerator [cf. Eq. (3) of the main text]: 

\begin{equation}
\overline{i {\bm f}(0,{\bm k},{\bm q}) \times {\bm f}^*(0,{\bm k},{\bm q})} 
= 
-\frac{4|\Delta|^2 {\cal A}^3\mu}{\Pi(0,k,0)^2}
\,
\overline{
{\bm k} \, 
[{\bm k} \cdot ({\bm z} \times {\bm q})]
}
=
\frac{2|\Delta|^2 {\cal A}^3 \mu}{\Pi(0,k,0)^2}\, k^2 ({\bm q} \times {\bm z}).
\label{ff*_Dirac_av}
\end{equation}
Here, the bar denotes the angle integral $\int_0^{2\pi}\frac{d\theta_{\bm k}}{2\pi}...$ over the directions of ${\bm k}$, and  
we used the identity $\overline{ k_a  k_b }=\frac{k^2}{2}\delta_{a,b}$ for the averaged product of 
the projections of ${\bm k}=[k_x, k_y, 0]$ (where $a,b = x,y$).

\section{\large B. Andreev conductance. Even and odd parts}

In this section, we calculate and analyze the zero-bias Andreev conductance in Eqs. (5) and (6) of the main text.  
We begin by rewriting the AR probability [Eq. (4) of the main text],
using the equations

\begin{equation}
\hat{F}_{_S}(E,{\bm k}) = \left[ f_0(E,{\bm k}) + {\bm \sigma} \cdot {\bm f}(E,{\bm k}) \right] i\sigma_y, \qquad
\hat{\rho}_{_F}(E,{\bm k})=\sum_{\alpha=\pm} \rho_\alpha (E,{\bm k}) \hat{P}_\alpha, \qquad 
\hat{P}_\alpha = \frac{1 + \alpha {\bm \sigma}\cdot {\bm m}}{2},
\nonumber
\end{equation}
as follows
\begin{eqnarray}
A(E) &=& \pi^2 \sum\limits_{ {\bm k}_1, {\bm k}_2, \alpha,\beta=\pm}
\sum_{{\bm k}{\bm k}^\prime} 
t_{{\bm k}_1,{\bm k}} t^*_{{\bm k}_2,{\bm k}} \left(t_{{\bm k}_1,{\bm k}^\prime} t^*_{{\bm k}_2,{\bm k}^\prime}\right)^*\times
\label{A}\\
&\times&
{\rm Tr}
\left[
\hat{P}_\alpha 
(f_0(E,{\bm k}) + {\bm \sigma} \cdot {\bm f}(E,{\bm k})) 
\mathbb{T} \hat{P}_\beta \mathbb{T}^{-1} 
(f^*_0(E,{\bm k}^\prime) + {\bm \sigma} \cdot {\bm f}^*(E,{\bm k}^\prime))
\right] 
\rho_\alpha(E,{\bm k}_1)\rho_\beta(-E,-{\bm k}_2).
\nonumber
\end{eqnarray}
Here, $\mathbb{T} \hat{P}_\beta \mathbb{T}^{-1}=\hat{P}_{-\beta}$ is the projector of the hole spin on the magnetization ${\bm m}$ in the F.
The hole spin state is obtained from the particle one by time-reversal operation $\mathbb{T}=i\sigma_y K$, where $K$ denotes complex conjugation. 
Equation (\ref{A}) can be written in the more compact form

\begin{eqnarray}
A(E) = \sum_{ {\bm k}_1, {\bm k}_2, \alpha,\beta=\pm}
{\rm Tr}\left[ \hat{r}^{\alpha,\beta}_{{\bm k}_1, {\bm k}_2}(E) \left[\hat{r}^{\alpha,\beta}_{{\bm k}_1, {\bm k}_2}(E)\right]^\dagger \right] 
\rho_\alpha(E,{\bm k}_1)\rho_\beta(-E,-{\bm k}_2),
\label{A3}
\end{eqnarray}
where $\hat{r}^{\alpha,\beta}_{{\bm k}_1, {\bm k}_2}(E)$ is a matrix in spin space given by
\begin{equation}
\hat{r}^{\alpha,\beta}_{{\bm k}_1, {\bm k}_2}(E) = \pi \sum\limits_{\bm k}
t_{{\bm k}_1,{\bm k}}t^*_{{\bm k}_2,{\bm k}}
\hat{P}_\alpha
[f_0(E,{\bm k}) + {\bm \sigma} \cdot {\bm f}(E,{\bm k})]
\mathbb{T} \hat{P}_\beta \mathbb{T}^{-1}.
\label{r}
\end{equation}
This matrix describes the conversion of the particle with spin $\alpha$ into the hole with spin $\beta$ 
through the creation of a Cooper pair in a mixed singlet-triplet state. 
Using the standard algebra of the Pauli matrices, we calculate the spin trace in Eq. (\ref{A3}): 
 
\begin{eqnarray}
{\rm Tr}\left[ \hat{r}^{\alpha,\beta}_{{\bm k}_1, {\bm k}_2}(E) \left[ \hat{r}^{\alpha,\beta}_{{\bm k}_1, {\bm k}_2}(E) \right]^\dagger \right]
&=& 
\pi^2 \Bigl\{
\Bigl|
\sum\limits_{\bm k} t_{{\bm k}_1,{\bm k}}
t^*_{{\bm k}_2,{\bm k}}
[
f_0(E,{\bm k}) + \alpha {\bm f}(E,{\bm k})\cdot {\bm m}
]
\Bigr|^2
\delta_{\beta,-\alpha}
\label{Tr_OS}\\
&+&
\frac{1}{2}
\Bigl\|
\sum\limits_{\bm k} t_{{\bm k}_1,{\bm k}}
t^*_{{\bm k}_2,{\bm k}}
\bigl[
{\bm m} \times ( {\bm f}(E,{\bm k}) \times {\bm m})-
i\alpha 
{\bm f}(E,{\bm k}) \times {\bm m}
\bigr]
\Bigr\|^2
\delta_{\beta,\alpha}
\Bigr\}.
\label{Tr_ES}
\end{eqnarray}
As mentioned in the main text, for non-collinear vectors ${\bm f}$ and ${\bm m}$ there are two different types of AR. 
The term in Eq. (\ref{Tr_OS}) corresponds to the process in which the spin of the reflected hole is opposite to that of the particle, $\beta=-\alpha$. 
Equation (\ref{Tr_ES}), on the contrary, describes the equal-spin AR in which the particle and the hole have parallel spin projections, $\beta=\alpha$. 
The equal-spin AR channel opens for a non-collinear orientation of the ${\bm f}$ and ${\bm m}$ vectors and 
is accompanied by the transfer of the spin angular momentum $\hbar$ and torques on the magnetization.    
As in the main text, the symbol $\|...\|$ above means the norm of a complex vector, e.g., $\|{\bm V}\| = \sqrt{ {\bm V}\cdot {\bm V}^*}$. 

We define the zero-bias and -temperature conductance as the derivative of the Andreev current, 
$G = \partial I_{_A}/\partial V|_{V,T \to 0}$. It is proportional to the AR probability taken at the Fermi level ($E=0$): $G = \frac{2e^2}{h} A(0)$. 
Using now Eqs. (\ref{A3}), (\ref{Tr_OS}), and (\ref{Tr_ES}) for the AR probability, 
we obtain the zero-bias Andreev conductance in Eq. (5) of the main text. 

Let us now derive the polarization-dependent part of the conductance, $G_{odd}$, Eq. (6) of the main text.
To that end, we change the order of the ${\bm k}$ summations in Eq. (5) of the main text, performing first the summations over ${\bm k}_1$ and ${\bm k}_2$: 

\begin{eqnarray}
G &=& \frac{2e^2}{h}\sum\limits_{{\bm k} {\bm k}^\prime \alpha=\pm }
\Gamma^\alpha_{ {\bm k},{\bm k}^\prime} \left(\Gamma^{-\alpha}_{ {\bm k},{\bm k}^\prime}\right)^* 
\left[
f_0(0,{\bm k}) + \alpha {\bm f}(0,{\bm k})\cdot {\bm m}
\right]
\left[
f^*_0(0,{\bm k}^\prime) + \alpha {\bm f}^*(0,{\bm k}^\prime)\cdot {\bm m}
\right]
\label{G_OS_1}\\
&+&
\frac{e^2}{h}\sum\limits_{{\bm k} {\bm k}^\prime \alpha=\pm }
\left|
\Gamma^\alpha_{ {\bm k},{\bm k}^\prime}
\right|^2
\left[
{\bm m} \times ({\bm f}(0,{\bm k}) \times {\bm m})
-i \alpha 
{\bm f}(0,{\bm k}) \times {\bm m}
\right]
\cdot
\left[
{\bm m} \times ({\bm f}^*(0,{\bm k}^\prime) \times {\bm m})
+ i\alpha 
{\bm f}^*(0,{\bm k}^\prime) \times {\bm m}
\right],
\label{G_ES_1}
\end{eqnarray}
where we introduced the energy scales

\begin{equation}
\Gamma^\alpha_{ {\bm k}, {\bm k}^\prime} = \pi \sum_{{\bm k}_1} t_{{\bm k}_1,{\bm k}}t^*_{{\bm k}_1,{\bm k}^\prime}\rho_\alpha(0,{\bm k}_1),
\label{Gamma}
\end{equation}
which determine the tunneling rates, $\Gamma^\alpha_{ {\bm k}, {\bm k}^\prime}/\hbar$, into the majority/minority F states. 
We also used the symmetry $\rho_\alpha(0,-{\bm k})=\rho_\alpha(0,{\bm k})$ of the F spectrum. Finally, we employ the algebraic identity

\begin{eqnarray}
&
\left[
{\bm m} \times ({\bm f}(0,{\bm k}) \times {\bm m})
-i \alpha 
{\bm f}(0,{\bm k}) \times {\bm m}
\right]
\cdot
\left[
{\bm m} \times ({\bm f}^*(0,{\bm k}^\prime) \times {\bm m})
+ i\alpha 
{\bm f}^*(0,{\bm k}^\prime) \times {\bm m}
\right] =
&
\label{Identity1}\\
&
2\left[
({\bm f}(0,{\bm k}) \times {\bm m}) \cdot ({\bm f}^*(0,{\bm k}^\prime) \times {\bm m}) + \alpha {\bm m} \cdot \left( i{\bm f}(0,{\bm k}) \times  {\bm f}^*(0,{\bm k}^\prime) \right)
\right],
&
\label{Identity2}
\end{eqnarray}
to cast the triplet conductance (\ref{G_ES_1}) into an alternative form:

\begin{eqnarray}
G &=& \frac{2e^2}{h}\sum\limits_{{\bm k} {\bm k}^\prime \alpha=\pm }
\Gamma^\alpha_{ {\bm k},{\bm k}^\prime} \left(\Gamma^{-\alpha}_{ {\bm k},{\bm k}^\prime}\right)^* 
\left[
f_0(0,{\bm k}) + \alpha {\bm f}(0,{\bm k})\cdot {\bm m}
\right]
\left[
f^*_0(0,{\bm k}^\prime) + \alpha {\bm f}^*(0,{\bm k}^\prime)\cdot {\bm m}
\right]
\label{G_OS_2}\\
&+&
\frac{2e^2}{h}\sum\limits_{{\bm k} {\bm k}^\prime \alpha=\pm }
\left|
\Gamma^\alpha_{ {\bm k},{\bm k}^\prime}
\right|^2
\left[
({\bm f}(0,{\bm k}) \times {\bm m}) \cdot ({\bm f}^*(0,{\bm k}^\prime) \times {\bm m}) + \alpha {\bm m} \cdot \left( i{\bm f}(0,{\bm k}) \times  {\bm f}^*(0,{\bm k}^\prime) \right)
\right].
\label{G_ES_2}
\end{eqnarray}
The last term in Eq. (\ref{G_ES_2}) accounts for the CSP associated with the non-unitary triplet pairing.
This terms changes the sign when either the pair polarization $i{\bm f}(0,{\bm k}) \times  {\bm f}^*(0,{\bm k}^\prime)$ or the F magnetization ${\bm m}$ changes its sign. 
That is why, in the main text and here, we use the notation $G_{odd}$ for this term. 
The rest of the conductance does not contain any terms like $i{\bm f}(0,{\bm k}) \times  {\bm f}^*(0,{\bm k}^\prime)$
and is even in ${\bm m}$ (see proof in Sec. C). 
It is convenient to separate the even and odd parts of Eqs. (\ref{G_OS_2}) and (\ref{G_ES_2}) as follows    

\begin{equation}
G = G_{even} + G_{odd},
\label{G_total}
\end{equation}
where 
\begin{eqnarray}
G_{even} &=& \frac{2e^2}{h}\sum\limits_{{\bm k} {\bm k}^\prime \alpha=\pm }
\Gamma^\alpha_{ {\bm k},{\bm k}^\prime} \left(\Gamma^{-\alpha}_{ {\bm k},{\bm k}^\prime}\right)^* 
\left[
f_0(0,{\bm k}) + \alpha {\bm f}(0,{\bm k})\cdot {\bm m}
\right]
\left[
f^*_0(0,{\bm k}^\prime) + \alpha {\bm f}^*(0,{\bm k}^\prime)\cdot {\bm m}
\right]
\label{G_even_s}\\
&+&
\frac{2e^2}{h}\sum\limits_{{\bm k} {\bm k}^\prime \alpha=\pm }
\left|
\Gamma^\alpha_{ {\bm k},{\bm k}^\prime}
\right|^2
({\bm f}(0,{\bm k}) \times {\bm m}) \cdot ({\bm f}^*(0,{\bm k}^\prime) \times {\bm m}).
\label{G_even_t}\\
G_{odd} &=& \frac{2e^2}{h}
\,
{\bm m} \cdot 
\sum_{ {\bm k}{\bm k}^\prime\alpha=\pm}
\alpha 
\left|
\Gamma^\alpha_{ {\bm k},{\bm k}^\prime}
\right|^2
i{\bm f}(0,{\bm k}) \times {\bm f}^*(0,{\bm k}^\prime).
\label{G_odd_1}
\end{eqnarray}

\section{\, \hskip 2cm \large C. Random tunneling model of the magnetoelectric Andreev effect}

In this section, we introduce the statistical description of Andreev tunneling in which the tunneling conductance
is treated as the average over the ensemble of random realizations of the tunneling matrix $t_{{\bm k},{\bm k}^\prime}$.    
Using this method, we will calculate the odd part of the conductance $G_{odd}$ [Eq. (7) of the main text] 
as well as the magnetoelectric Andreev conductance in Eq. (8) of the main text. 

\subsection{\large C1. Correlation function of $t_{{\bm k},{\bm k}^\prime}$ and averaging procedure}

We describe the tunneling by the Hamiltonian 
\begin{equation}
H_T =
\sum_{
s {\bm k} {\bm k^\prime} 
} 
\left[ b^\dagger_{s {\bm k}} \, t_{{\bm k},{\bm k^\prime} } \, a_{s {\bm k^\prime} }
+  
a^\dagger_{s {\bm k^\prime} } \, t^*_{ {\bm k},{\bm k^\prime} } \, b_{s {\bm k}} 
\right],
\label{H_T}
\end{equation}
which introduces the hybridization of the electronic states of the "left" and "right" systems (represented by the operators $a_{s {\bm k}}$ and $b_{s {\bm k}}$, respectively). 
Since the tunneling Hamiltonian (\ref{H_T}) is quadratic, it can always be mapped to the Hamiltonian of the electron-impurity problem. 
We can therefore assume that the hybridization occurs at local impurity sites at the interface between the "left" and "right" systems. 
In this case, $t_{{\bm k},{\bm k^\prime} }$ can be treated as the Fourier transform of the real-space potential of the interface impurities. 
We assume further that the impurity ensemble is random and, by the central limiting theorem, $t_{{\bm k},{\bm k^\prime} }$ is a Gaussian-distributed random variable.  
Its statistical properties are fully characterized by a two-point correlation function which can be defined by analogy with the random impurity problem: 
\begin{equation}
\langle\langle t_{{\bm k}_1,{\bm k} } t_{{\bm k}_2,{\bm k^\prime} } \rangle\rangle = \zeta_{{\bm k}_1-{\bm k}_2}\delta_{{\bm k}_1-{\bm k}, -{\bm k}_2 +{\bm k^\prime}},
\label{corr}
\end{equation}
where $\langle\langle ... \rangle\rangle$ denotes averaging over the random positions of the interface impurities and 
$\zeta_{|{\bm k}_1-{\bm k}_2|}$ is the correlation function in momentum space. 
This is a well-defined minimal model for disordered interfaces that occur in a number of experimental situation, e.g., in contacts created by the sputtering technique.

We identify the observable tunneling conductance with the average over the random positions of the interface impurities. 
The averaging of the Andreev conductance in Eqs. (\ref{G_OS_2}) and (\ref{G_ES_2}) reduces to the calculation of the average product of the tunneling rates,  

\begin{equation}
\langle\langle
\Gamma^\alpha_{ {\bm k},{\bm k}^\prime} \left(\Gamma^\beta_{ {\bm k},{\bm k}^\prime}\right)^*
\rangle\rangle 
=\pi^2 \sum_{{\bm k}_1,{\bm k}_2} 
\langle\langle
t_{{\bm k}_1,{\bm k}}t^*_{{\bm k}_1,{\bm k}^\prime} t^*_{{\bm k}_2,{\bm k}}t_{{\bm k}_2,{\bm k}^\prime}
\rangle\rangle 
\rho_\alpha(0,{\bm k}_1)\rho_\beta(0,{\bm k}_2).
\label{GamGam_av}
\end{equation}
To proceed further it is necessary to eliminate the complex conjugated matrix elements. This can be done by using the TRS relations $t^*_{{\bm k}_1,{\bm k}^\prime}=t_{-{\bm k}_1,-{\bm k}^\prime}$ and 
$t^*_{{\bm k}_2,{\bm k}}=t_{-{\bm k}_2,-{\bm k}}$. Then, the average of the four Gaussian-distributed matrix elements 
falls into products of the correlators (\ref{corr}) through the pair contractions:

\begin{eqnarray}
\langle\langle
t_{{\bm k}_1,{\bm k}} t^*_{{\bm k}_1,{\bm k}^\prime} t^*_{{\bm k}_2,{\bm k}} t_{{\bm k}_2,{\bm k}^\prime}
\rangle\rangle 
=
\langle\langle
t_{{\bm k}_1,{\bm k}} t_{-{\bm k}_1,-{\bm k}^\prime} t_{-{\bm k}_2,-{\bm k}} t_{{\bm k}_2,{\bm k}^\prime}
\rangle\rangle 
=
\label{tttt}
\end{eqnarray}
\begin{eqnarray} 
\langle\langle
t_{{\bm k}_1,{\bm k}} t_{-{\bm k}_1,-{\bm k}^\prime} 
\rangle\rangle 
\langle\langle
t_{-{\bm k}_2,-{\bm k}} t_{{\bm k}_2,{\bm k}^\prime}
\rangle\rangle 
+
\langle\langle
t_{{\bm k}_1,{\bm k}} t_{{\bm k}_2,{\bm k}^\prime} 
\rangle\rangle 
\langle\langle
t_{-{\bm k}_1,-{\bm k}^\prime}  t_{-{\bm k}_2,-{\bm k}}
\rangle\rangle 
+
\langle\langle
t_{{\bm k}_1,{\bm k}} t_{-{\bm k}_2,-{\bm k}} 
\rangle\rangle 
\langle\langle
t_{-{\bm k}_1,-{\bm k}^\prime} t_{{\bm k}_2,{\bm k}^\prime}
\rangle\rangle =
\label{tt_tt}
\end{eqnarray}
\begin{eqnarray} 
\zeta_{{\bm k}_1-{\bm k}} \delta_{{\bm k}, {\bm k}^\prime} \,\, \zeta_{{\bm k}-{\bm k}_2} \delta_{{\bm k}, {\bm k}^\prime} 
+
\zeta_{{\bm k}_1-{\bm k}} \delta_{{\bm k}_1+{\bm k}_2, {\bm k}+{\bm k}^\prime} \,\, \zeta_{{\bm k}^\prime - {\bm k}_1} \delta_{{\bm k}_1+{\bm k}_2, {\bm k}+{\bm k}^\prime}
+
\zeta_{{\bm k}_1-{\bm k}} \delta_{{\bm k}_1, {\bm k}_2} \,\, \zeta_{{\bm k}^\prime-{\bm k}_1} \delta_{{\bm k}_1, {\bm k}_2}.
&
\label{z_z} 
\end{eqnarray}  
In the following, we restrict ourselves to the short-ranged real-space correlations which in ${\bm k}$ space have a constant correlation function $\zeta_{\bm k}\equiv \zeta = const$. 
In this case, Eq. (\ref{z_z}) is reduced to 

\begin{eqnarray}
\langle\langle
t_{{\bm k}_1,{\bm k}} t^*_{{\bm k}_1,{\bm k}^\prime} t^*_{{\bm k}_2,{\bm k}} t_{{\bm k}_2,{\bm k}^\prime}
\rangle\rangle 
= \zeta^2 ( \delta_{{\bm k}, {\bm k}^\prime} + \delta_{{\bm k}_1+{\bm k}_2, {\bm k}+{\bm k}^\prime} + \delta_{{\bm k}_1, {\bm k}_2}).
\label{z_z_short} 
\end{eqnarray} 
Inserting this equation into Eq. (\ref{GamGam_av}) yields

\begin{eqnarray}
\langle\langle
\Gamma^\alpha_{ {\bm k},{\bm k}^\prime}
\left(\Gamma^\beta_{ {\bm k},{\bm k}^\prime}\right)^*
 \rangle\rangle 
&=&
\pi^2 \zeta^2 \delta_{{\bm k}, {\bm k}^\prime} 
\left( \sum_{{\bm k}_1} \rho_\alpha(0,{\bm k}_1)\right) \left( \sum_{{\bm k}_2} \rho_\beta(0,{\bm k}_2)\right)+
\label{GamGam_av_lead}\\
&+&
\pi^2 \zeta^2 
\sum_{{\bm k}_1} 
\rho_\alpha(0,{\bm k}_1)\rho_\beta(0,{\bm k}+{\bm k}^\prime - {\bm k}_1) 
+
\pi^2 \zeta^2 \sum_{{\bm k}_1} \rho_\alpha(0,{\bm k}_1)\rho_\beta(0,{\bm k}_1). 
\label{GamGam_av_rest}
\end{eqnarray}
Here, the first term involves the double sum over ${\bm k}$, whereas the rest are single series over ${\bm k}$. 
Due to this fact, the first term provides the leading contribution to the average conductance. 
In the following, we will only keep the leading contribution (\ref{GamGam_av_lead}).   

\subsection{\large C2. Odd, even and relative magnetoelectric Andreev conductances}

We start with the calculation of the average odd conductance from Eq. (\ref{G_odd_1}).
With the help of Eq. (\ref{GamGam_av_lead}) the averaging of Eq. (\ref{G_odd_1}) yields 

\begin{eqnarray}
G_{odd} = 
\frac{2e^2}{h}
\,
{\bm m} \cdot 
\sum_{ {\bm k}{\bm k}^\prime\alpha=\pm}
\alpha 
\langle\langle
\left|
\Gamma^\alpha_{ {\bm k},{\bm k}^\prime}
\right|^2
 \rangle\rangle 
i{\bm f}(0,{\bm k}) \times {\bm f}^*(0,{\bm k}^\prime)
&=&
\frac{2e^2}{h}
(\Gamma^2_+ - \Gamma^2_-)
\,
{\bm m} \cdot 
\sum_{\bm k}
i{\bm f}(0,{\bm k}) \times {\bm f}^*(0,{\bm k}),
\label{G_odd_2}\\
\Gamma_\alpha = \pi \zeta \rho_\alpha, 
\qquad 
\rho_\alpha &=& \sum_{\bm k} \rho_\alpha(0,{\bm k}).
\label{Gamma_DOS}
\end{eqnarray}
Here, $\Gamma_\alpha$ are the tunneling rates for a disordered interface. They are proportional to the densities of states at the Fermi level, $\rho_\alpha$, and 
the correlator of the tunneling matrix, $\zeta$. 
This result is quoted in Eq. (7) of the main text and is valid for a generic non-unitary pairing. 
Let us now consider specifically the linear Dirac surface state with the supercurrent-induced CSP [see Eqs. (\ref{Pi_Dirac}) 
and (\ref{ff*_Dirac_av})]. In this case, the ${\bm k}$ integration in Eq. (\ref{G_odd_2}) can be done exactly:

\begin{eqnarray}
G_{odd}({\bm q}) &=& 
\frac{2e^2}{h}
(\Gamma^2_+ - \Gamma^2_-)
\,
a\int_0^\infty \frac{kdk}{2\pi} 
\,
{\bm m} \cdot 
\left(
\overline{
i{\bm f}(0,{\bm k},{\bm q}) \times {\bm f}^*(0,{\bm k},{\bm q})
}
\right)
\label{G_odd_Dirac1}\\
&=&
{\bm m} \cdot ({\bm q} \times {\bm z}) \,
\frac{2e^2}{h}
(\Gamma^2_+ - \Gamma^2_-)
\,
\frac{|\Delta|^2 {\cal A}^3 \mu a}
{\pi}
\int_0^\infty \frac{k^3dk}
{\Pi(0,k,0)^2}
\label{G_odd_Dirac2}\\
&=&
\frac{e^2}{h} \,
(\Gamma^2_+ - \Gamma^2_-) \,
\frac{
{\bm z} \cdot ({\bm m} \times {\bm q}) a
}
{
2\pi\mu {\cal A}
}
\,
{\cal F}_t \left( \frac{\Delta}{\mu}\right).
\label{G_odd_Dirac3}
\end{eqnarray}
Here, $a$ is the surface area and ${\cal F}_t(x)$ is the dimensionless function of the ratio $x= \Delta/\mu$ given by

\begin{equation}
{\cal F}_{t}(x)= 1 +\frac{1 - x^2}{2x}\left( \frac{\pi}{2} + \arctan\frac{1-x^2}{2x} \right).
\label{F_t}
\end{equation}

In Eq. (\ref{G_odd_Dirac3}), the odd ${\bm q}$ dependence $G_{odd}$ is explicit. 
This part of the conductance is also odd in the magnetization ${\bm m}$. Interestingly, 
since the phase gradient ${\bm q}$ is parallel to the surface, Eq. (\ref{G_odd_Dirac3}) depends 
only on the in-plane component ${\bm m}_\|$ of the total magnetization ${\bm m}$. 
That is, if the F is magnetized perpendicularly to the surface, there is no odd contribution to the total Andreev conductance 
for the disordered interface.   

In the same manner, we obtain the average even part of the conductance from Eqs. (\ref{G_even_s}) and (\ref{G_even_t}): 

\begin{eqnarray}
G_{even} &=& 
\frac{2e^2}{h}
\sum\limits_{{\bm k} \alpha=\pm }
\Gamma_\alpha 
\Gamma_{-\alpha}
\left|
f_0(0,{\bm k}) + \alpha {\bm f}(0,{\bm k}) \cdot {\bm m}
\right|^2
+
\frac{2e^2}{h}
\sum\limits_{{\bm k} \alpha=\pm }
\Gamma^2_\alpha
\left\|
{\bm f}(0,{\bm k}) \times {\bm m}
\right\|^2
\label{G_even_1}\\
&=&
\frac{4e^2}{h}\Gamma_+\Gamma_-  
\sum_{\bm k}
\left(
\left|
f_0(0,{\bm k})
\right|^2
+
\left|
{\bm f}(0,{\bm k}) \cdot {\bm m}
\right|^2
\right)
+
\frac{2e^2}{h}(\Gamma^2_+ + \Gamma^2_-)
\sum_{\bm k}
\left\|
{\bm f}(0,{\bm k}) \times {\bm m}
\right\|^2.
\label{G_even_2}
\end{eqnarray}
In Eq. (\ref{G_even_2}), we summed up the spin terms, taking into account the symmetry of $\Gamma_\alpha\Gamma_{-\alpha}$ in the first (opposite-spin) term.
Not only is $G_{even}$ symmetric under ${\bm m}\to - {\bm m}$, but it also does not change under any of the transformations 

\begin{equation}
f_0(0,{\bm k}) \to -f_0(0,{\bm k}), \qquad {\rm Re}{\bm f}(0,{\bm k}) \to - {\rm Re}{\bm f}(0,{\bm k}), \quad {\rm and} \quad {\rm Im}{\bm f}(0,{\bm k}) \to - {\rm Im}{\bm f}(0,{\bm k}).
\label{transformations}
\end{equation}
In particular, for the supercurrent-induced nonunitary state, $G_{even}$ is a symmetric function of the phase gradient ${\bm q}$ independently of the details 
of $f_0(0,{\bm k},{\bm q})$ and ${\bm f}(0,{\bm k},{\bm q})$ in Eqs. (\ref{f_s}) and (\ref{f_t}). 
Therefore, the total Andreev conductance is generically asymmetric under the reversal of the supercurrent flow:

\begin{equation}
G({\bm q}) - G(-{\bm q}) = 2G_{odd}({\bm q}) \qquad {\rm or} \qquad G({\bm j}) - G(-{\bm j}) = 2G_{odd}({\bm j})
\label{G_q}
\end{equation}   
where phase gradient ${\bm q}$ is related to the supercurrent density ${\bm j}=\nu_{_S}\nabla\varphi=2\nu_{_S} {\bm q}$. 
Within the linear response in ${\bm j}$, the relative change in the conductance is 

\begin{equation}
\frac{G({\bm j}) - G(-{\bm j})}{G({\bm j})} \approx \frac{2G_{odd}({\bm j})}{G(0)} =  \frac{2G_{odd}({\bm j})}{G_{even}(0)},
\label{dG_rel}
\end{equation} 
where $G_{odd}({\bm j})$ is given by Eq. (\ref{G_odd_Dirac3}) and the zero-current conductance $G_{even}(0)$ acts as the normalization constant. 
The latter can be calculated from Eqs. (\ref{G_even_2}), (\ref{f_s}) and (\ref{f_t}) with ${\bm \Omega}_{\bm k} = {\cal A} {\bm k} \times {\bm z}$ and $\epsilon_{\bm k}=0$,
using the same integration procedure as in Eqs. (\ref{G_odd_Dirac1}) - (\ref{G_odd_Dirac3}):  

\begin{equation}
G_{even}(0)=
\frac{e^2}{h} \,
\Gamma_+\Gamma_- \,
\frac{
a
}
{
2\pi {\cal A}^2
}
\,
{\cal F}_s \left( \frac{\Delta}{\mu}\right) 
+
\frac{e^2}{h} \,
\left[
\left(
\frac{
\Gamma_+ + \Gamma_-
}{2}
\right)^2
+
\left(
\frac{
\Gamma_+ - \Gamma_-
}{2}
\right)^2
{\bm m}^2_\perp
\right]
\frac{
a
}
{
2\pi {\cal A}^2
}
\,
{\cal F}_t \left( \frac{\Delta}{\mu}\right), 
\label{G_even_0}
\end{equation}
where ${\bm m}_\perp$ is the out-of-plane magnetization component and ${\cal F}_{s}(x)$ is the dimensionless function 

\begin{equation}
{\cal F}_{s}(x)= 1 +\frac{1 + x^2}{2x}\left( \frac{\pi}{2} + \arctan\frac{1-x^2}{2x} \right),
\label{F_s}
\end{equation}
which comes from the $k$-integration in the singlet term. 
We note that $G_{even}(0)$ depends on the orientation of the magnetization ${\bm m}$ with respect to the surface. 
This result agrees with the prediction of Ref. \cite{Hoegl15} which considered an S/F junction with a Rashba SO-coupled interface 
in the absence of the supercurrent. 

From Eqs. (\ref{G_odd_Dirac3}) and (\ref{G_even_0}) 
we find the magnetoelectric Andreev conductance as

\begin{equation}
\frac{G({\bm j}) - G(-{\bm j})}{G(0)} = \frac{2G_{odd}({\bm j})}{G_{even}(0)} = 
{\cal A} \frac{{\bm z} \cdot ( {\bm m} \times {\bm j} )}{2\nu_{_S}\mu} \, 
\frac{
4(\Gamma^2_+ - \Gamma^2_-)
}
{
\left(
\Gamma_+ + \Gamma_-
\right)^2
+
\left(
\Gamma_+ - \Gamma_-
\right)^2
{\bm m}^2_\perp
+ 
4 \Gamma_+\Gamma_- {\cal F}_s\left(\frac{\Delta}{\mu}\right)/{\cal F}_t \left( \frac{\Delta}{\mu}\right)}.
\label{dG_rel_1}
\end{equation}
Since $\Gamma_\pm =\pi \zeta \rho_\pm$ depend on the densities of the majority and minority states at the Fermi level, $\rho_\pm$, 
we can write the switching conductance in terms of the F spin polarization $P = \frac{ \rho_+ - \rho_- }{\rho_+ + \rho_-}$, 
as we did in Eq. (8) of the main text.

\end{document}